\documentclass[a4paper,11pt]{article}
\usepackage{pos}
\usepackage{amsmath}
\usepackage{textcomp, gensymb}
\usepackage{graphicx}

\usepackage{physics}
\usepackage{xcolor}

\title{Analytical description of the radio Cherenkov cone}
\ShortTitle{Radio Cherenkov description}

\author*[a]{Valentin Decoene}
\author[b,c]{Marion Guelfand}
\author[d]{Mat\`ias Tueros}

\affiliation[a]{SUBATECH, Institut Mines-Telecom Atlantique, CNRS/IN2P3, Université de Nantes, Nantes, France}

\affiliation[b]{Sorbonne Université, CNRS, Laboratoire de Physique Nucléaire et de Hautes Energies (LPNHE), Paris, France}

\affiliation[c]{Institut d’Astrophysique de Paris, CNRS, Sorbonne Université, Paris, France}

\affiliation[d]{IFLP - CCT La Plata - CONICET, Depto. de Física, Fac. de Cs. Ex., Universidad Nacional de La Plata, La Plata, Argentina}

\emailAdd{valentin.decoene@subatech.in2P3.fr}
\emailAdd{marion.guelfand@lpnhe.in2p3.fr}

\abstract{Radio emissions from extensive air showers (EAS) provide a valuable tool for detecting ultra-high-energy (UHE) astroparticles. In this context, several radio arrays focus on detecting highly inclined EAS, as this enables the observation of Earth-skimming UHE neutrinos, in addition to cosmic rays and gamma rays.
The reconstruction of such inclined events relies heavily on a thorough understanding of the radio features observed on the ground, with the Cherenkov cone being one of the most prominent. In this study, we demonstrate that the Cherenkov cone can be accurately described for inclined air showers using basic propagation principles. Furthermore, we have developed an analytical model that computes the expected opening angles of the cone and reproduces the asymmetry effects observed in simulations. The high accuracy of these computations can enhance current reconstruction methods and pave the way for the development of new ones.}

\ConferenceLogo{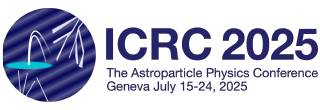}

\FullConference{39th International Cosmic Ray Conference (ICRC2025)\\
 15–24 July 2025\\
Geneva, Switzerland\\}

\begin{document}
\maketitle

\section{Introduction}
\vspace{-0.3cm}
Radio detection of ultra-high-energy (UHE) astroparticles, especially cosmic rays, has proven to be effective~\cite{Lecoz_2017,Abreu_2012,Huege_2008,ARDOUIN_2005} and is now being explored for detecting earth-skimming UHE neutrinos in projects like BEACON~\cite{Southall_2023}, GRAND~\cite{GRAND_WP}, and PUEO~\cite{PUEO:2023zrz}. 
Understanding radio emision from the extensive air showers (EAS) that
these UHE neutrinos and cosmic rays induce in the atmosphere is
crucial
for accurately reconstructing the event's direction, energy, and nature. A key feature of the radio signal is the so-called Cherenkov cone, where the signal is strongly enhanced, making it easier to detect and important to understand for accurate predictions.

Radio signals from air showers result from a combination of emission mechanisms and propagation effects. Microscopically, the particle front generates radiation from billions of charged particles, each emitting in a well-defined beam pattern described by classical electrodynamics. The lobes of this pattern have an opening angle given by $\alpha=\arccos{\qty(1/n)}$, where $n$ is the refractive index. Macroscopically, this radiation travels through the atmosphere to the observer. When $n > 1$, particles in the shower front move faster than light in air, causing time compression effects. These effects lead to the simultaneous arrival of emissions from different points along the shower, resulting in a signal enhancement at specific locations, forming the Cherenkov cone (see e.g.,~\cite{deVries_2011,AlvarezMuniz_2010,Alvarez_muniz_2012,CORSTANJE_2017}). The refractive index’s variation with altitude complicates this for inclined showers, where the interplay of refractive index, Earth's curvature, shower altitude, and radiation path variations causes asymmetries that have been observed in simulations but that
have not been accurately modeled yet~\cite{Guelfand_2025,Schluter_2020}.

In Sec.~\ref{sec:analytical_model}, we present the analytical framework developed to model the Cherenkov effect and its angles for inclined air showers, and in Sec.~\ref{sec:simulation_comparison}, we compare these predictions to simulations.

\vspace{-0.3cm}
\section{Analytical model for the Cherenkov cone}\label{sec:analytical_model}
\begin{figure}
    \centering
    \includegraphics[width=0.7\linewidth]{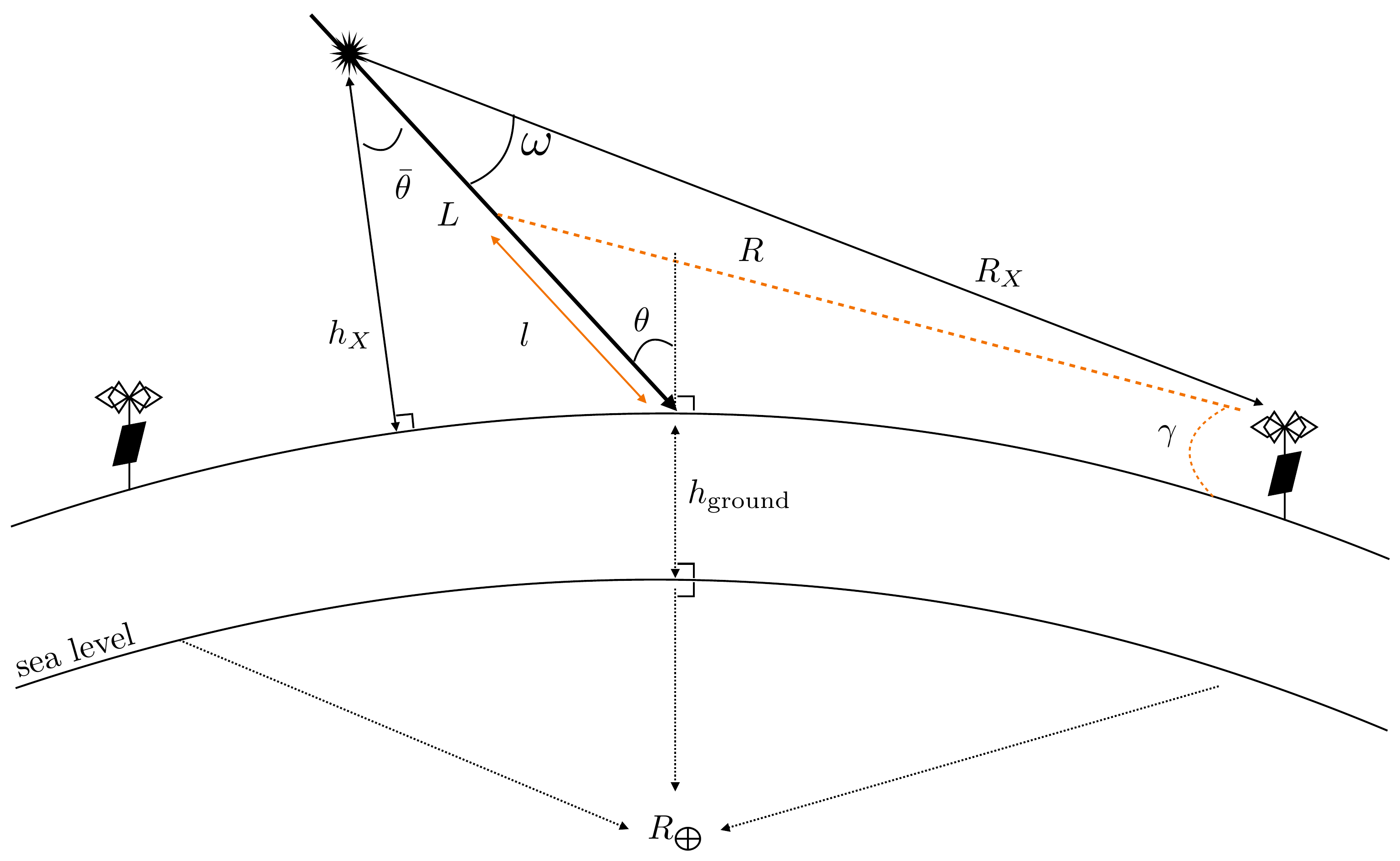}
    \caption{The sketch represents the geometry, accounting for Earth's curvature. The shower hits the ground with a local zenith angle $\theta$, and the $X_{\rm max}$ location is marked as a black spiky sphere along the shower axis, with altitude $h_X$ zenith angle $\bar{\theta}$, and distance $L$ to the core. The observer (antenna) is defined by the angle $\omega$ to the shower axis and the distance $R_X$ to $X_{\rm max}$. Emissions along the shower axis are characterized by their longitudinal position $l$, distance $R$, and viewing angle $\gamma$ to the observer.}
    \label{fig:sketch_curved_ground}
\end{figure}

\vspace{-0.3cm}
Our hypothesis, based on previous studies, is the interaction between the $X_{\rm max}$ location (where particle number peaks) and the time compression (boost) effect at specific observer locations. In Sec.~\ref{sec:refrac_index}, we compute the refractive index, which is then used in Sec.~\ref{sec:boost_factor} to calculate the time compression and associated Cherenkov angles. The air-shower geometry is described using angular coordinates (Fig.~\ref{fig:sketch_curved_ground}), and the Earth’s curvature is taken into consideration to accurately model very inclined showers.

\vspace{-0.2cm}
\subsection{Index of refraction} \label{sec:refrac_index}
\begin{figure}
    \centering
\includegraphics[width=0.49\linewidth]{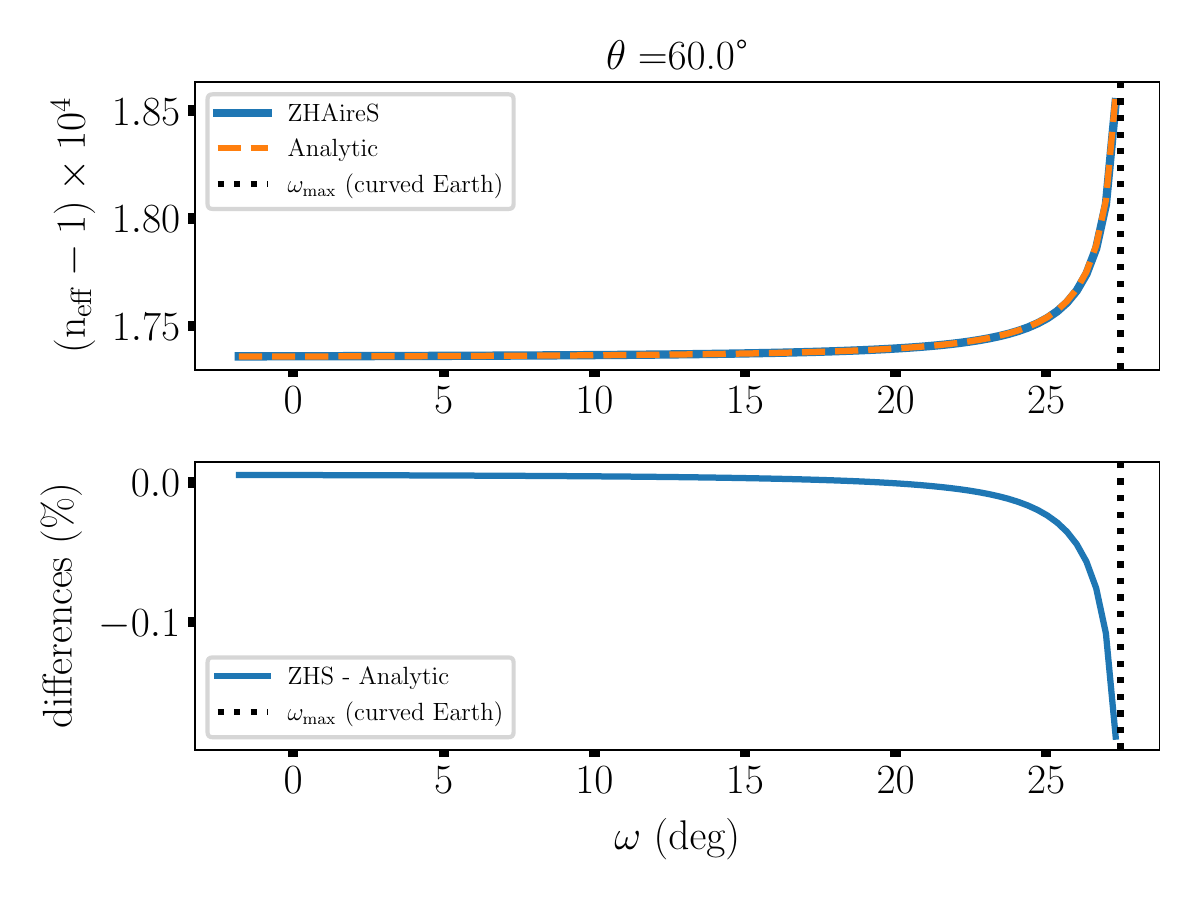}
\includegraphics[width=0.49\linewidth]{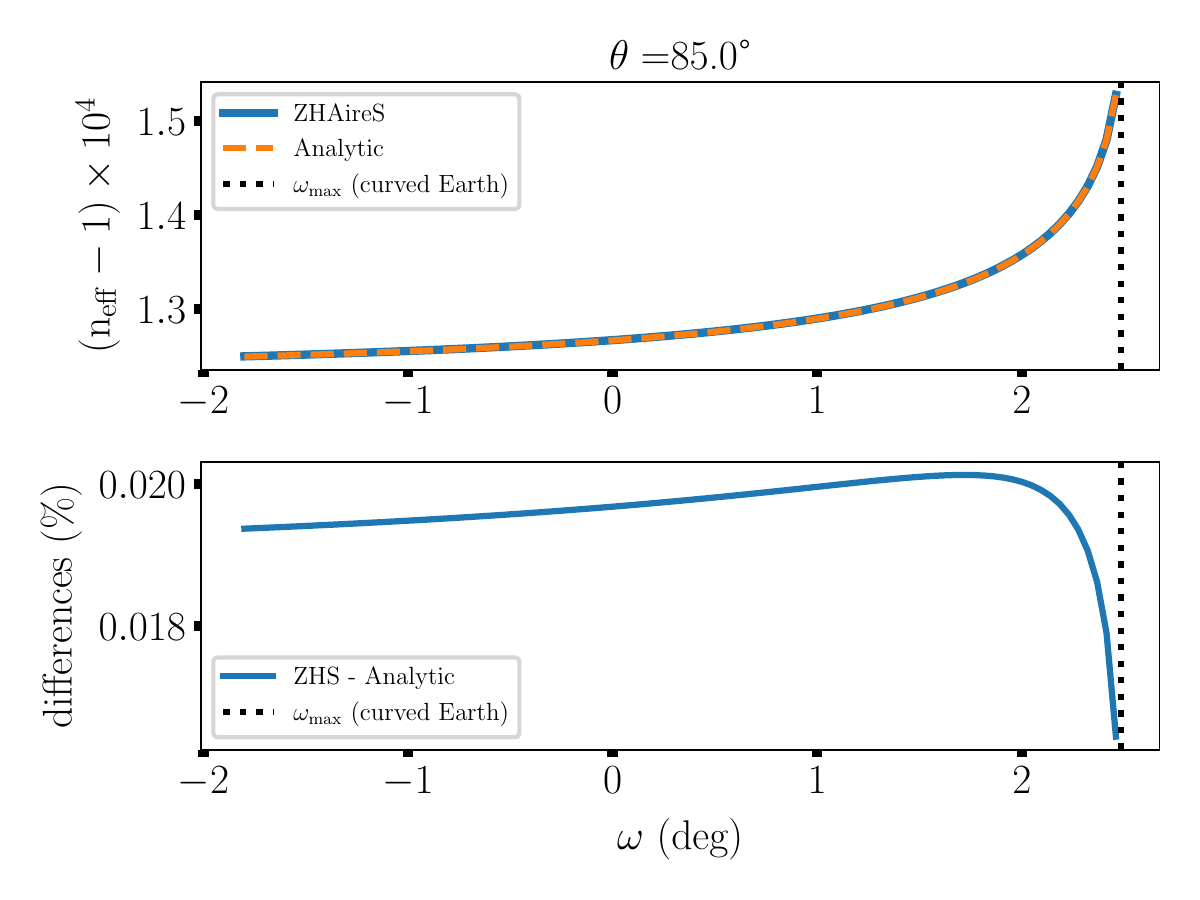}
    \caption{Effective refractive index vs. observer angle $\omega$ for the ZHAireS algorithm (blue solid line) and analytical calculation (dashed orange line), shown for zenith angles of $60\degree$ ({\it left}) and $85\degree$ ({\it right}). {\it Bottom panels} display the relative difference between both methods. The vertical dotted black line marks the maximum observer angle allowed by Earth’s curvature.}
    \label{fig:refrac_comparisons}
\end{figure}

\vspace{-0.2cm}
Modeling the index of refraction dependence with altitude is central
to our aproach. Since in this work we will be comparing with results
from state-of-the art EAS simulations made with ZHAireS~\cite{Alvarez_muniz_2012,Sciutto_2019}, we will
focus on its single-exponential model: $n\qty(h) = 1 + k e^{-C h}$, where $h$ is the altitude, $k= 3.25.10^{-4}$ and $C= 0.1218$\,km$^{-1}$.
The average refractive index, along the path $R$ between the emission point and the observer, is defined by
\begin{align}
    \label{eq:effective_refractive_index}
    \expval{n\qty(R)} = \frac{\int_0^R \dd{r} n\qty(r)}{\int_0^R \dd{r}} \ ,
\end{align}
where $r$ is the integration variable along the traveled path $R$, assumed to be a straight line. This expression therefore reduces to
\begin{align} \label{eq:effective_refractive_index_I}
    \expval{n\qty(R)} &= 1 + \frac{k}{R} \int_0^R e^{-C h\qty(r)} \dd{r} \equiv 1 + \frac{k}{R} I\qty(R) \ ,
\end{align}
where the term $I\qty(R)$ is detailed in App.~\ref{sec:appendix_refractive_index}. 
Fig.~\ref{fig:refrac_comparisons} compares the effective refractive index from ZHAireS (that computes the integral numerically) with our analytical model (Eq.~\ref{eq:effective_refractive_index_I}) for shower inclinations of $\theta=60\degree$ and $85\degree$, across observer angles $\omega$ from $-2\degree$ to $1.1\times \omega_{\rm max}$ (the maximum visible angle due to Earth’s curvature; see App.~\ref{sec:appendix_geometry}). The bottom panels show the relative error, which stays below $\sim 0.2\%$ even for steep showers.

\vspace{-0.2cm}
\subsection{Boost factor and Cherenkov angles} \label{sec:boost_factor}
\vspace{-0.2cm}
The time compression effect arises from the causal link between emission time $t'$ and observer time $t$, related by the optical path:
$c\qty(t-t') = \expval{n\qty(R)} R$, 
where $R$ is the distance between emission and observer, and $\expval{n\qty(R)}$ is the effective refractive index along the path (see Sec.~\ref{sec:refrac_index}). With a realistic, altitude-dependent refractive index, the $t'$-to-$t$ relation becomes complex and can be multi-valued, meaning multiple emission times correspond to one observer time. This critical point, influenced by shower inclination and observer position, defines the emission boost (if emissions are occurring at the corresponding altitude).

The boost factor is defined where the emitted signal along the shower axis is observed over a very short time, formally where $\dv{ct'}{ct} \to \infty$. For convenience, we use its inverse, $\dv{ct}{ct'} \to 0$, since zeros are easier to handle. Using the retarded time and effective refractive index (Eq.~\ref{eq:effective_refractive_index_I}), the boost factor is expressed as:
\begin{align} \label{eq:general_boost_exp}
    \dv{ct}{ct'} = 1 + \qty[\dv{\expval{n}}{R} R + \expval{n}] \dv{R}{ct'} = 1 + \qty[1+ k\dv{I\qty(R)}{R}] \frac{ct' - R_X \cos{\qty(\omega)}}{R} \ ,
\end{align}
with $\dv{R}{ct'} = \frac{ct' - R_X \cos{\qty(\omega)}}{R}$ (see App.~\ref{sec:appendix_geometry} and ~\ref{sec:appendix_cherenkov_angle} for details). A boosting effect occurs wherever this multi-valued function reaches zero and it is observed only if particles emit at that particular time. The interplay between the particle emission distribution and the causal link between retarded and observer times —shaped by the effective refractive index— produces the radio Cherenkov effect and its characteristic Cherenkov cone (or angles).

\begin{figure}[ht]
    \centering
\includegraphics[width=0.99\linewidth]{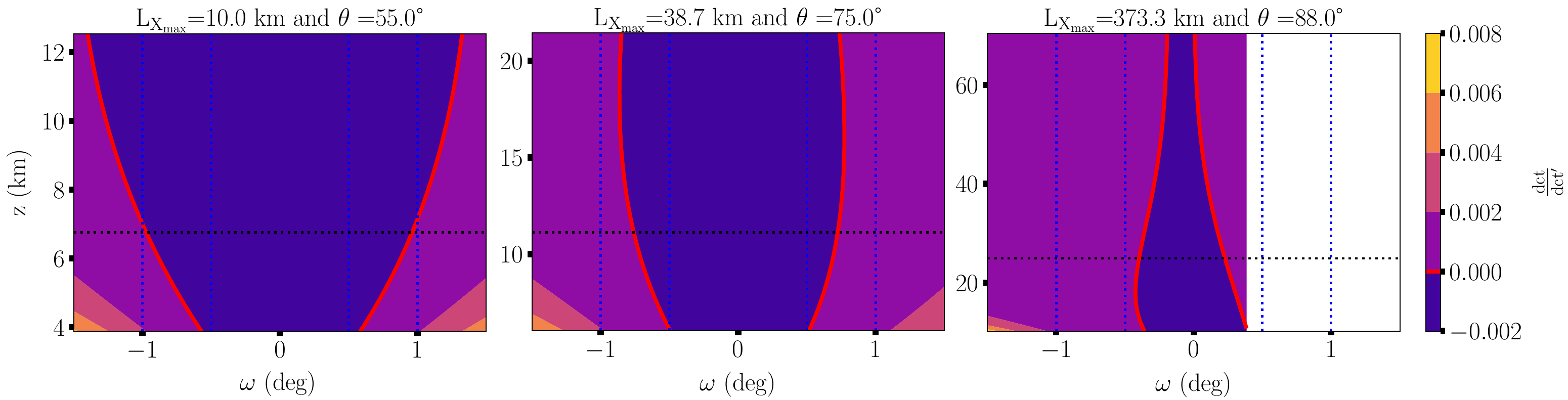}
    \caption{Inverse boost factor vs. altitude $z$ and viewing angle $\omega$ for shower directions of $55\degree$, $75\degree$, and $88\degree$. The expected $X_{\rm max}$ altitude is shown as a dotted horizontal black line. The red line marks where the inverse boost factor drops to zero (i.e., infinite boost). In the {\it right panel}, the white area indicates regions where the boost factor is undefined due to Earth’s curvature blocking the line of sight.}
    \label{fig:boost_map}
\end{figure}

The boosting effect defines the Cherenkov cone angle $\omega_{\rm Ch}$, where $\dv{ct}{ct'}\eval_{\omega = \omega_{\rm Ch}} = 0$.
Fig.\ref{fig:boost_map} shows this derivative’s variation with shower altitude $z$ and observer angle $\omega$ for different inclinations. The red line marks where the derivative cancels, causing boosting. While multiple shower locations can boost, the Cherenkov cone is assumed to form near the maximum particle emission ($X_{\rm max}$), corresponding to the observer angle at that boosted altitude (with retarded time fixed to $X_{\rm max}$, i.e., $ct'=0$). Changes in $X_{\rm max}$ altitude and boosting with inclination cause the Cherenkov angle $\omega_{\rm Ch}$ to vary.
Such effect has been observed in simulations, showing a shift in the Cherenkov angle between early and late observers, which grows with increasing shower inclination (see e.g.,~\cite{Guelfand_2025,Schluter_2020}).
Interestingly, if we set the refractive index to a constant $\expval{n}=n$, Eq.~\ref{eq:general_boost_exp} becomes
\begin{align}
    \dv{ct}{ct'} = 1 + n \frac{ct' - R_X \cos{\qty(\omega)}}{R} \ ,
\end{align}
and from our definition of the Cherenkov angle $\dv{ct}{ct'} = 0$ at $X_{\rm max}$ (which implies $ct'=0$ and $R=R_X$), we find the well known "standard" formula of the Cherenkov angle $\omega_{\rm Ch} = \arccos{\qty(1/n)}$.

\vspace{-0.3cm}
\section{Comparison with simulations} \label{sec:simulation_comparison}
\vspace{-0.3cm}

\vspace{-0.1cm}
\subsection{Simulation setup} \label{sec:simulation_setup}
\vspace{-0.2cm}
We simulate extensive air showers (EAS) using AIRES v19.04.08 with the Sibyll 2.3d hadronic model and a thinning level of $10^{-5}$. The resulting electric fields are computed using ZHAireS v1.0.30a, accounting for contributions from all electrons and positrons. Atmospheric density follows the extended Linsley model, and the refractive index varies with altitude following the exponential profile in Sec.~\ref{sec:refrac_index}, with a sea-level index of $1.000325$ and ground altitude set at $1086$\,m.

The dataset includes 1700 proton and iron showers with azimuths of 0° or 180° (North–South axis). Zenith angles follow a logarithmic distribution in $1/\rm cos(\theta)$ and energies range from 0.02 to 3.98 EeV in 22 logarithmic bins —covering the range detectable by radio arrays. 

The antenna array uses a star-shaped layout centered on the shower core, with 152 observers placed in 20 angular directions per arm, covering $\omega \leq 3\degree$.

\vspace{-0.2cm}
\subsection{Results} \label{sec:results}
\vspace{-0.2cm}
The Cherenkov angle extraction from simulations proceeds as follows: we first identify the signal peak at each antenna using the maximum of the Hilbert envelope of the electric field norm. For each of the 8 array arms, we select the antenna with the highest peak. The angular distance between each selected antenna and the shower axis (measured from the simulated $X_{\rm max}$ position) defines the Cherenkov angle $\omega_{\rm Ch}$. The antenna spacing of $0.15\degree$ per arm introduces a limit on angular precision.

\begin{figure}[ht]
    \centering
     \includegraphics[width=0.64\linewidth]{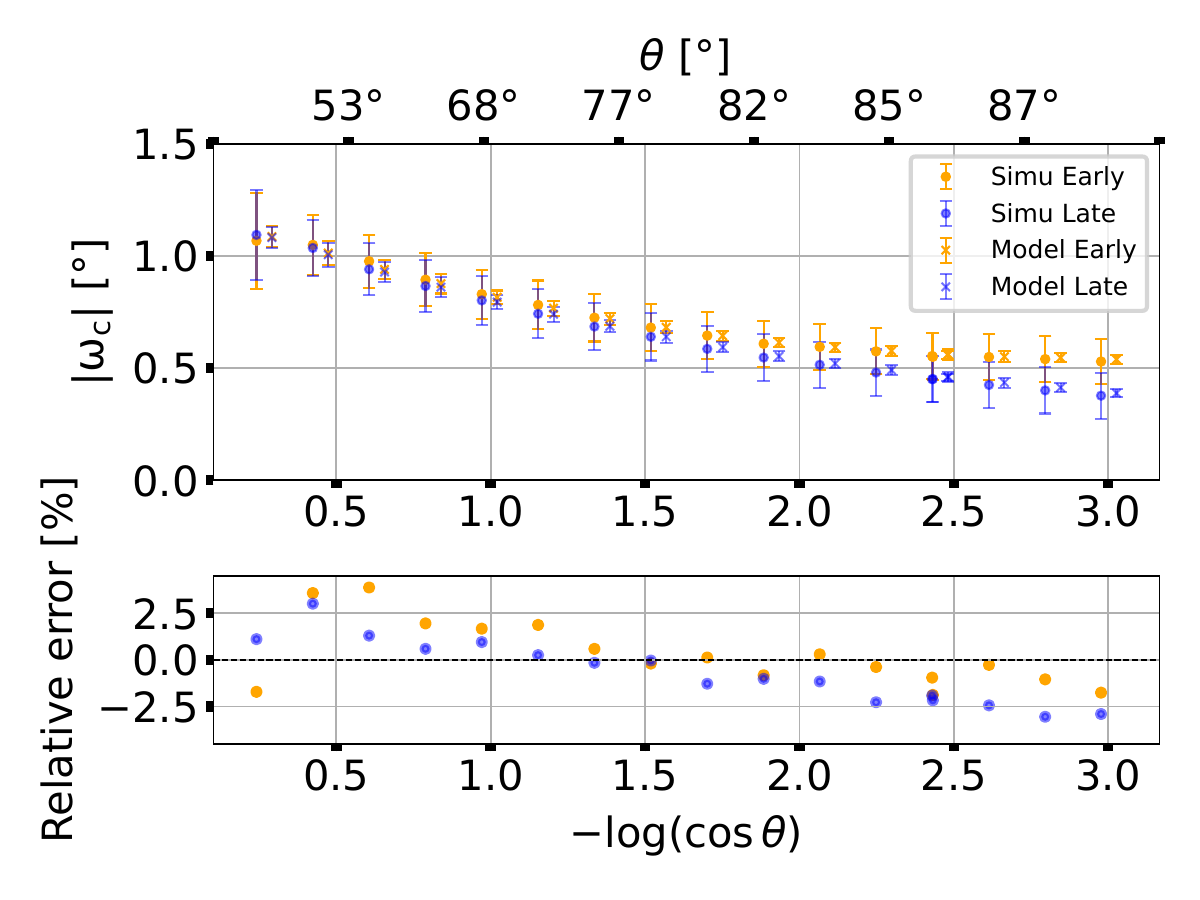}
     \\\includegraphics[width=0.64\linewidth]{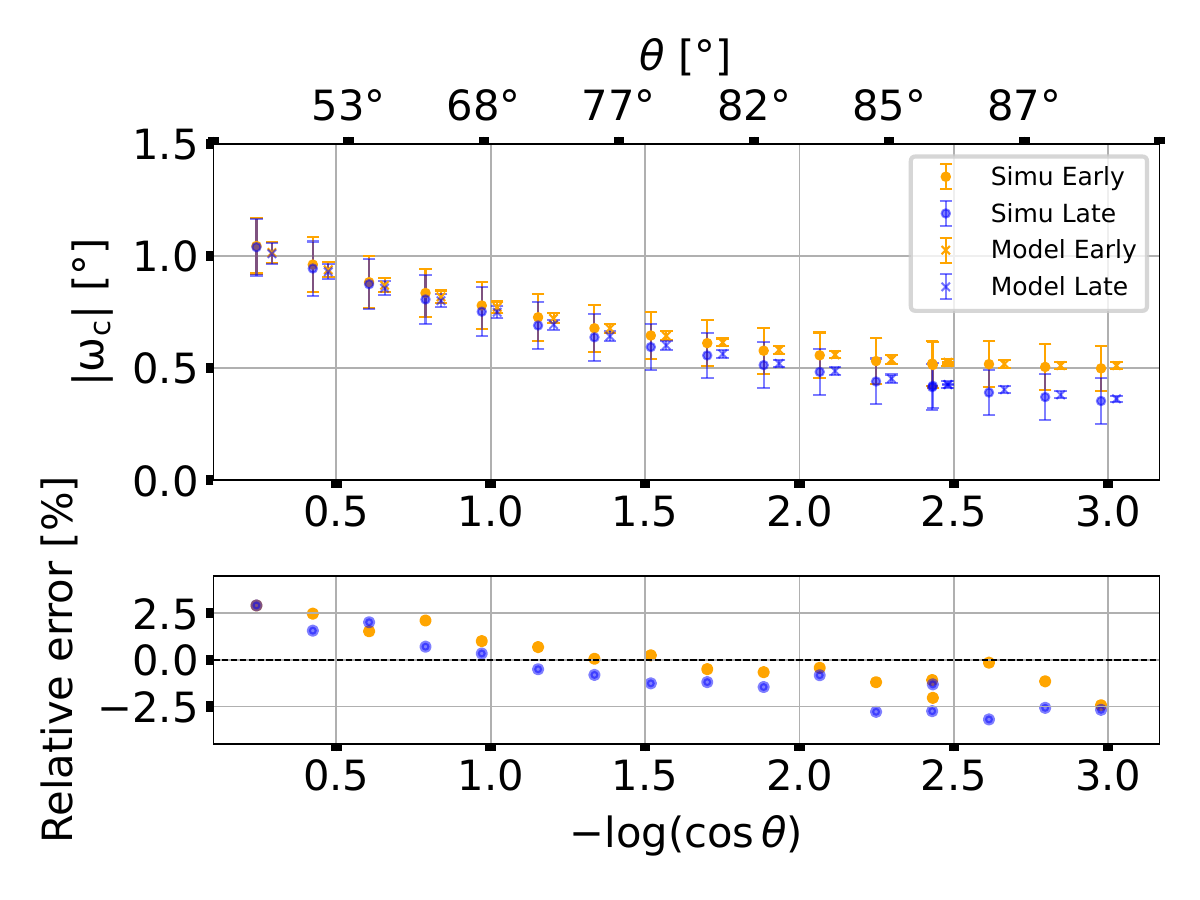}
    \caption{
    Evolution of $\omega_c$ with shower inclination along the propagation direction. Dots show simulation results; crosses indicate analytical values. Orange markers correspond to early antennas (closer to emission), blue to late antennas (see Fig.~\ref{fig:sketch_curved_ground}). Error bars combine statistical and systematic uncertainties, the latter reflecting the angular step size —larger in simulations. The {\it top panel} shows proton results; the {\it bottom panel}, iron.}
    \label{fig:cherenkov_angles_results}
\end{figure}

Fig.~\ref{fig:cherenkov_angles_results} compares simulated $\omega_{\rm Ch}$ values (full frequency range) with analytical predictions versus shower inclination. For each event, we show only two $\omega_{\rm Ch}$ values from the arms aligned with the propagation direction. The analytical model agrees very well with simulations across all zenith angles, including the asymmetry at high inclinations, where two distinct $\omega_{\rm Ch}$ values emerge.

\vspace{-0.3cm}
\section{Conclusion}
\vspace{-0.3cm}
In this study, we developed a model of the radio Cherenkov effect based on fundamental propagation principles. The model focuses on variations in optical paths along the air-shower trajectory and through the atmosphere, presented here in a full geometrical framework accounting for ground curvature and inclined showers. This allows for an analytical calculation of Cherenkov angles that accurately reproduces simulations, with angular precision better than $0.02\degree$ ($2.5\%$ in relative error), capturing both the overall trend and asymmetry. The model also yields analytical expressions for the effective refractive index.

Both could be useful for reconstruction methods such as in the Angular Distribution Function (ADF~\cite{Guelfand_2025,Guelfand_procICRC}), which estimates shower geometry and energy from observer angle-dependent amplitudes, and therefore relies on the Cherenkov angles and effective refractive index computations. 

Additionally, the model could help reconstruct the $X_{\rm max}$ position. Since determining Cherenkov angles requires $X_{\rm max}$, the model can be inverted to estimate $X_{\rm max}$ from measured angles at antennas. While systematic effects (e.g., angle reconstruction, shower direction, atmospheric refractivity) must be evaluated, the method could at least offer a useful cross-check or complementary estimate of $X_{\rm max}$ and atmospheric grammage.

Finally, the formalism presented here will be extended to multi-layered exponential atmospheric models in a future publication.

\appendix

\vspace{-0.3cm}
\section{Geometrical computations} \label{sec:appendix_geometry}
\vspace{-0.3cm}
The expression for the local altitude along the line of sight $h\qty(r)$ (where $r$ is ranging from $0$ to $R$) is given by solving the triangle $\qty(h\qty(r)+R_{\bigoplus}, r, R_{\bigoplus})$. Which gives
\begin{align} \label{eq:hlocal}
    h\qty(r) = - R_{\bigoplus} + \sqrt{r^2 + R_{\bigoplus}^2 - 2r R_{\bigoplus} \cos{\qty(\gamma)}} \ .
\end{align}
From the triangle $\qty(h_X + R_{\bigoplus}, L_X, R_{\bigoplus})$ we obtain the local altitude for the $X_{\rm max}$ location
\begin{align}
    h_X = -R_{\bigoplus} + \sqrt{L_X^2+R_{\bigoplus}^2 + 2L_X R_{\bigoplus}\cos{\qty(\theta)}} \ ,
\end{align}
with $L_X$ being the longitudinal distance between $X_{\rm max}$ and the ground.
In the triangle $\qty(h_X + R_{\bigoplus}, R_X, R_{\bigoplus})$ we derive the distance between an observer (located in the direction defined by $\omega$ w.r.t the shower axis) and $X_{\rm max}$
\begin{align}
   R_X = \qty(h_X + R_{\bigoplus})\cos{\qty(\bar{\theta}+\omega)} \pm \sqrt{R_{\bigoplus}^2 - \qty(h_X + R_{\bigoplus})^2 \sin^2{\qty(\bar{\theta}+\omega)}} \ ,
\end{align}
after solving the second order polynomial in $R_X$, and where $\bar{\theta}$ is the local zenith angle of the shower.
Using the triangle $\qty(ct', R_X, R)$ we have the distance between an emission point (located at $ct'$ along the shower axis) and an observer
\begin{align}
    R = \sqrt{\qty(ct')^2 + R_X^2 - 2ct' R_X \cos{\qty(\omega)}} \ .
\end{align}
The local height $h_0$ of an emission point along the shower axis is defined from the triangle $\qty(h_0+R_{\bigoplus}, L_X -ct', R_{\bigoplus})$, which gives
\begin{align}
    h_0 = - R_{\bigoplus} + \sqrt{\qty(L_X - ct')^2 + R_{\bigoplus}^2 + 2\qty(L_X - ct')R_{\bigoplus} \cos{\qty(\theta)}} \ .
\end{align}
The angle $\gamma$ defining the direction of the line of sight from the observer location is obtained from the triangle $\qty(h_0+R_{\bigoplus}, R, R_{\bigoplus})$
\begin{align}
   \cos{\qty(\gamma)} = \frac{R^2 - h_0^2 - 2h_0 R_{\bigoplus}}{2 R R_{\bigoplus}} \ .
\end{align}
The angle $\bar{\theta}$ defining the local zenith angle can be obtained from the triangle $\qty(R_{\bigoplus}, L_X, h_X + R_{\bigoplus})$
\begin{align}
   \cos{\qty(\bar{\theta})} = \frac{L_X + R_{\bigoplus} \cos{\qty(\theta)}}{h_X + R_{\bigoplus}} \ .
\end{align}
In the case of a curved ground, there is a maximal value for $\omega$ beyond which no light path can be drawn from the $X_{\rm max}$ point and the observer, because the observer is below the horizon
\begin{align}
    \omega_{\rm max} = \arcsin{\qty(\frac{R_{\bigoplus}}{h_X + R_{\bigoplus}})} - \bar{\theta} \ ,
\end{align}
this is also true for any emission point at an altitude below $X_{\rm max}$ (obtained by replacing $h_X$ by $h_0$).

\vspace{-0.3cm}
\section{Refractive index computations} \label{sec:appendix_refractive_index}
\vspace{-0.3cm}
To integrate the effective refractive index as defined in Sec.~\ref{sec:refrac_index}, we develop $h\qty(r)$ (Eq.~\ref{eq:hlocal}) at the second order in terms of $r/R_{\bigoplus}$, which gives
\begin{align}
    h\qty(r) \approx -r \cos{\qty(\gamma)} + \frac{1}{2}\frac{r^2}{R_{\bigoplus}}\qty(1 -{\cos^2{\qty(\gamma)}}) + \mathcal{O}\qty(\frac{r^3}{R_{\bigoplus}^3}) \ .
\end{align}
Injecting the above expression into the effective refractive index expression (Eq.~\ref{eq:effective_refractive_index}), leads to
\begin{align}
      \expval{n\qty(R)} &= 1 + \frac{k}{R} \int_0^R e^{-C h\qty(r)} \dd{r} =  1 + \frac{k}{R} \int_0^R e^{C\qty(r\cos{\qty(\gamma) - \frac{r^2}{2R_{\bigoplus}}\qty[1 - \cos^2{\qty(\gamma)}]})} \dd{r} \ .
\end{align}
Where the integral part, noted $I\qty(R)$ can be express as
\begin{align}
   I\qty(R) =  \sqrt{\frac{\pi R_{\bigoplus}}{2C \Gamma}}\ e^{\frac{R_{\bigoplus}C \cos^2{\qty(\gamma)}}{2 \Gamma}} \qty[\erf{\qty(\sqrt{\frac{R_{\bigoplus}C}{2}\Gamma}\ \qty(\frac{R}{R_{\bigoplus}} - \frac{\cos{\qty(\gamma)}}{\Gamma}))} + \erf{\qty(\sqrt{\frac{R_{\bigoplus}C \cos^2{\qty(\gamma)}}{2\Gamma}})}] \ ,
\end{align}
with $\Gamma=1 - \cos^2{\qty(\gamma)}$.

\vspace{-0.3cm}
\section{Cherenkov angles computations} \label{sec:appendix_cherenkov_angle}
\vspace{-0.3cm}
The computation of the Cherenkov angle as defined in Sec.~\ref{sec:boost_factor} (see Eq.~\ref{eq:general_boost_exp}), requires the derivative of $I\qty(R)$, given by
\begin{align}
    &\dv{I\qty(R)}{R} = I\qty(R) \frac{\cos{\qty(\gamma)}}{\Gamma} \dot{\gamma} \qty(1+\frac{CR_{\bigoplus}}{\Gamma})+ e^{CR\qty(\cos{\qty(\gamma)} - \frac{\Gamma R}{2 R_{\bigoplus}})} \qty[1 - \frac{\dot{\gamma}}{\Gamma}\qty(\frac{R_{\bigoplus}}{\Gamma} + \cos{\qty(\gamma)}R)] + \frac{ R_{\bigoplus}}{\Gamma^2} \dot{\gamma} \ ,
  \\ &{\rm with\ }  \dot{\gamma} = \dv{\cos{\qty(\gamma)}}{R} = \frac{1}{R_{\bigoplus}} - \frac{\cos{\qty(\gamma)}}{R} + \qty(\frac{h_0}{R_{\bigoplus}}+1)\frac{L_X - ct' + R_{\bigoplus}\cos{\qty(\theta)}}{\qty[ct'-R_X\cos{\qty(\omega)}]\qty[h_0 + R_{\bigoplus}]} \ .
\end{align}

\vspace{-0.3cm}
\bibliographystyle{ICRC}
\setlength{\bibsep}{0pt plus 0.3ex}
\bibliography{biblio}

\providecommand{\href}[2]{#2}\begingroup\raggedright\begin{thebibliography}{10}

\bibitem{Lecoz_2017}
S.~Le~Coz {\em et~al.} \href{http://dx.doi.org/10.22323/1.301.0406}{{\em
  ICRC2017} (July, 2017) 406}.

\bibitem{Abreu_2012}
{The Pierre Auger {\it Coll.}}
  \href{http://dx.doi.org/10.1088/1748-0221/7/10/P10011}{{\em JINST} {\bfseries
  7} no.~10, (2012) P10011}.

\bibitem{Huege_2008}
T.~Huege {\em et~al.}
  \href{http://dx.doi.org/10.1088/1742-6596/110/6/062012}{{\em J. Phys: Conf.
  Ser.} {\bfseries 110} no.~6, (May, 2008) 062012}.

\bibitem{ARDOUIN_2005}
D.~Ardouin {\em et~al.}
  \href{http://dx.doi.org/10.1016/j.nima.2005.08.096}{{\em NIM-A} {\bfseries
  555} no.~1, (2005) 148--163}.

\bibitem{Southall_2023}
D.~Southall {\em et~al.}
  \href{http://dx.doi.org/10.1016/j.nima.2022.167889}{{\em NIM-A} {\bfseries
  1048} (2023) 167889}.

\bibitem{GRAND_WP}
{GRAND {\it Coll.}} \href{http://dx.doi.org/10.1007/s11433-018-9385-7}{{\em
  Sci. China Phys. Mech. Astron.} {\bfseries 63} no.~1, (2020) 219501}.

\bibitem{PUEO:2023zrz}
{PUEO {\it Coll.}} \href{http://dx.doi.org/10.22323/1.444.1028}{{\em PoS}
  {\bfseries ICRC2023} (2023) 1028}.

\bibitem{deVries_2011}
K.~D. {de Vries} {\em et~al.}
  \href{http://dx.doi.org/10.1103/PhysRevLett.107.061101}{{\em Phys. Rev.
  Lett.} {\bfseries 107} no.~6, (Aug., 2011) 061101}.

\bibitem{AlvarezMuniz_2010}
J.~{Alvarez-Mu{\~n}iz} {\em et~al.}
  \href{http://dx.doi.org/10.1103/PhysRevD.81.123009}{{\em Phys. Rev. D}
  {\bfseries 81} no.~12, (June, 2010) 123009}.

\bibitem{Alvarez_muniz_2012}
J.~Alvarez-Muñiz {\em et~al.}
  \href{http://dx.doi.org/https://doi.org/10.1016/j.astropartphys.2011.10.005}{{\em
  Astroparticle Physics} {\bfseries 35} no.~6, (2012) 325--341}.

\bibitem{CORSTANJE_2017}
A.~Corstanje {\em et~al.}
  \href{http://dx.doi.org/https://doi.org/10.1016/j.astropartphys.2017.01.009}{{\em
  Astroparticle Physics} {\bfseries 89} (2017) 23--29}.

\bibitem{Guelfand_2025}
M.~Guelfand {\em et~al.}
  \href{http://dx.doi.org/https://doi.org/10.1016/j.astropartphys.2025.103120}{{\em
  Astroparticle Physics} {\bfseries 171} (2025) 103120}.

\bibitem{Schluter_2020}
F.~Schl\"uter {\em et~al.}
  \href{http://dx.doi.org/10.1140/epjc/s10052-020-8216-z}{{\em Eur. Phys. J. C}
  {\bfseries 80} no.~7, (2020) 643}.

\bibitem{Sciutto_2019}
{S. J. Sciutto}, ``{AIRES} a system for air shower simulations. user's guide
  and reference manual,'' 2019.

\bibitem{Guelfand_procICRC}
M.~Guelfand {\em et~al.} {\em PoS} {\bfseries ICRC2025} (these proceedings)
  278.

\end{thebibliography}\endgroup

\end{document}